\documentclass[epj,nopacs]{svjour}
\usepackage{latexsym}
\usepackage{graphics}
\usepackage{hyperref}

\begin{document}
\title{Moffat's Modified Gravity tested on X-COP galaxy clusters}
\subtitle{}
\author{Sreekanth Harikumar\inst{1} \and Marek Biesiada \inst{1}
}                     
\offprints{}          
\institute{ National Centre for Nuclear Research, Pasteura 7,  Warsaw 02-093, Poland}
%
\date{Received: date / Revised version: date}
%
\abstract{
Scalar Tensor Vector Gravity (STVG) is a fully covariant  Lorentz invariant alternative theory of gravity also known as MOdified Gravity (MOG) which modifies General Relativity by inclusion of dynamical massive vector field and scalar fields. In STVG the mass $\mu$ of the vector field $\phi$ and the   gravitational constant G acquire the status of dynamical fields. We use the reconstructed total cluster mass of the X-COP sample obtained from X-ray observations by XMM-Newton telescope in combination with  Sunyaev-Zel'dovich (SZ) effect observed within Planck all-sky survey to estimate the $ \alpha$ and $\mu$ parameters of MOG theory. The obtained values are consistent with previous fits by other authors. Hence the MOG is passing another test and proves its consistency strengthening thereby its stance of being a promising alternative to General Relativity.
\PACS{
      {PACS-key}{discribing text of that key}   \and
      {PACS-key}{discribing text of that key}
     } 
} 
\maketitle
\section{Introduction}
\label{intro}
A century after Albert Einstein developed General  (GR) we made the first direct detection of gravitational waves \cite{Abbott_2016} confirming the validity of Einstein’s equations in the strong and highly dynamical regime. This detection can also be called a milestone in fundamental physics as it is one of the direct tests of GR confirming its validity. General Relativity has passed all tests in the solar system and binary pulsar scales and it has now become an unavoidable tool for astrophysicists. During this period we have also witnessed a rise of many modified gravity theories as an alternative to the GR and over the years one have witnessed considerable blows in experimental tests. The recent resurgence in exploring and proposing new theories were driven by the need to address the issues like dark energy, dark matter and inflation apart from the difficulties in developing a quantum theory of gravity. The nature of dark matter is one of major mysteries of contemporary physics. It began in the early 1930's when \cite{Zwicky:1933gu} reported an excessive rotational velocity of the luminous matter present in galaxies. It gained popularity with Vera Rubin's observation that the outskirts of spiral galaxies are rotating faster than they should.
There are several approaches to explain this missing mass in the universe. The most common approach is that dark matter constituents could be Weakly Interacting Massive Particle (WIMPs) or Massive Astrophysical Compact Halo Objects (MACHOs). Despite several experimental efforts \cite{Liu_2017} to find these particle no evidence in favour of them \cite{Tan_2016} has been reported. Another possible candidate for dark matter particle is axion which is an  extremely light bosonic particle that is expected to be detected in ground based experiments and so far there has been no conclusive evidence for its existence \cite{KIM19871}. In order to remedy this situation many people proposed alternatives to GR which solve the problem of missing mass in a natural way without the need for any dark component.

The period 1960-1980 can be called the Golden Era in experimental gravitational physics during which there were numerous efforts to verify the predictions of GR. At the heart of GR is the equivalence principle which underlies metric description of gravity. 
Any competing theory should obey Einstein's equivalence principle \cite{1993tegp.book.....W} which states that matter should couple universally to a single  metric tensor field, a symmetric second rank tensor $g_{\alpha\beta}$. Such class of theories can be called metric theories of gravity. There could be other dynamical or non-dynamical gravitational fields present in the spacetime besides the metric but they are prevented from interacting with matter. The factor that distinguishes one metric theory from the other is the number of gravitational fields it contains in addition to the metric.  
Alternative theories of gravity usually have more degrees of freedom in comparison to the GR. Significant deviations from GR can be currently probed by searching for extra polarization in the gravitational waves. Alternative theories have more polarizations besides plus and cross as a consequence of these extra degrees of freedom. With the increasing sensitivity of the upgraded ground based GW detectors and planned space borne projects like LISA, DECIGO or TianQin we will be able to detect signatures of modified gravity beyond the GR, if such possibility is realized in Nature.

Moffat’s theory of gravity \cite{Moffat_2006} is a covariant modification of GR  introducing two more fundamental fields (responsible for gravity) besides the metric tensor: a scalar field and a vector field. Hence it is sometimes called Scalar-Tensor- Vector Gravity (STVG).
In this theory, the gravitational constant $G$ and the mass of vector field $\mu$ are not constant but are allowed to vary with space and time. In the weak field limit Moffat’s gravity reduces to ordinary Newtonian potential along with a Yukawa like repulsive force \cite{MoffatRahvar2014}. Unlike the relativistic version of Milgrom's MOND \cite{Milgrom} in which the speeds of light and gravitons are different \cite{Green_2018}, STVG has photons and GWs that follow the same null geodesics of the respective spacetime. The STVG has so far demonstrated good fits to the galaxy rotation curves   
and cluster data 
without non- baryonic matter or compatibility with CBM cosmological data \cite{MoffatRahvar2014,rotationcurves2013,zahra2021,MoffatandToth2015,GREEN2019,moffat2020}.
Thus, the inability of GR alone in explaining the ``dark'' components of the Universe is overcome naturally.
Galaxy clusters are the most massive astrophysical laboratories which account for a larger fraction of mass content in the Universe. It is believed that $80 \% $ of the matter content in these clusters is in the form of dark matter interacting only through gravity. Galaxies that we observe in the electromagnetic spectrum including their stellar components and the intergalactic gas (shining in X-rays) account of the remaining $20 \% $ matter present. In this paper we use the X-COP data of 12 nearby massive galaxy clusters for which we have high-confidence total mass estimates to estimate the weak field parameters in MOG acceleration law. The paper has been organised as follows: In Section 2 we present the  key concepts of STVG theory. The  modified acceleration law in the weak field limit of it is also discussed. In Section 3 we derive the modified cluster mass in STVG assuming hydrostatic equilibrium and estimate the parameters in two different scenario using the X-COP cluster data. Last section summarized the results.

\section{Weak field limit of Scalar Tensor Vector Gravity}

Moffat's modified gravity theory postulates the existence of a massive vector field $\phi$ of mass $\mu$, which introduces a repulsive modification of the law of gravitation at short range. 
The vector field is coupled universally to matter. 
The theory promotes $G$ and $\mu$ to scalar fields, hence they are allowed to run, resulting in the following action 
\cite{Moffat_2006},\cite{rotationcurves2013}

\begin{equation}
S =  S_G + S_{\phi} + S_{s} + S_{M}
\end{equation}
where $S_M$ is the matter action,
\begin{equation}
    S_{G}  = \frac{1}{16\pi}\int d^{4}x \sqrt{-g} \frac{1}{G}\left( R + 2 \Lambda \right)  
\end{equation}
is the usual Einstein-Hilbert action for gravity,
\begin{equation}
S_{\phi} = \int d^{4}x \sqrt{-g}\left[ -\frac{1}{16\pi}B^{\alpha\beta}B_{\alpha\beta} + \frac{1}{8\pi}\mu^{2}\phi_{\alpha}\phi^{\alpha} - 
V_{\phi}(\phi_{\alpha} \phi^{\alpha})\right]
\end{equation}
is the action for the vector field $\phi$, with $B_{\alpha \beta} = \partial_{\alpha} \phi_{\beta} - \partial_{\beta} \phi_{\alpha}$ being the Faraday tensor for the vector field and
\begin{eqnarray}
&&S_{s} = \int d^{4}x\sqrt{-g} \Big[ \frac{1}{G^{3}}\left( \frac{1}{2}g^{\alpha\beta}\nabla_{\alpha}G\nabla_{\beta}G - V_{G}(G) \right) \\ 
&& + \frac{1}{\mu^{2}G}\Big( \frac{1}{2}g^{\alpha\beta}\nabla_{\alpha}\mu\nabla_{\beta}\mu - V_{\mu}({\mu}) \Big) \Big]   \nonumber
\end{eqnarray}
is the action for the scalar fields $G$ and $\mu$, $V_{\phi}(\phi)$, $V_{G}$ and $V_{\mu}(\mu)$ denote self-interaction potentials for the vector field and scalar fields. 

On astrophysical scales, studying the behaviour of
MOG, one can use the weak field approximation for the dynamics of
gravitating systems with perturbing them around Minkowski space
for the arbitrary distribution of non-relativistic matter. Under this assumption, it has been shown in \cite{Moffat_2006,MoffatandToth2015} that the scalar fields remain constant and the 
Newtonian law is modified with a Yukawa like repulsive term in the presence of vector field $\phi$ of mass $\mu$. Linearised field equations and corresponding gravitational potential have been derived in \cite{rotationcurves2013}. 
\begin{equation}
    \Phi_{eff}(\vec{x})  =  -G_{N}\int \frac{\rho(\vec{x^{'}})}{|\vec{x}-\vec{x}^{'}|}\left (\ 1+\alpha -\alpha e^{-\mu |\vec{x}-\vec{x}^{'}|}\right )\ d^{3}x^{'}
\end{equation}
where $G_{N}$ is the Newton's gravitational constant and $\alpha$ is defined through the field $G = G_{N}(1+\alpha)$. The parameter  $\alpha$ is a dimensionless number which measures the strength of the  attractive part of the potential.
Let us recall, that $Q_g=\sqrt{\alpha G_N}M$ is the gravitational charge 
 of the vector field $\phi_{\mu}$ and plays an important role for applications of the Schwarzschild-MOG static 
 spherically symmetric solution (MOG black holes and other fits to data) \cite{MOG_BH}. The effective acceleration of a test body can be obtained using $a_{eff} = - \nabla \Phi_{eff}$. For a point mass we obtain the modified acceleration law in the form:
\begin{equation}\label{modifiedacceleration}
    a_{eff}    = \frac{G_{N}M}{r^{2}}\left (\  1+\alpha-\alpha e^{-\mu r}(1+\mu r)\right )\
\end{equation}
The above acceleration reduces to the acceleration in a Newtonian theory when $\alpha=0$. The mass of the vector field $\mu$  is treated as an effective running parameter and it scales with the mass of the  physical system. Its evolution is governed by  \cite{Moffat:2016lkt}
\begin{equation}
  \Box \mu = \frac{1}{G}\partial^{\alpha}G\partial_{\alpha}\mu + \frac{2}{\mu}\partial^{\alpha}\mu\partial_{\alpha}\mu  +  \mu^{2}G\frac{V(\phi)}{\partial \mu}
\end{equation}
Previously, estimates of the two parameters in the modified acceleration have been obtained for galaxy rotation curves and few other astrophysical scenarios. One particular choice \cite{rotationcurves2013} of the parameters that fits the Milky Way's rotation curve is $\alpha_{MW} = 8.89$ and $\mu_{MW} = 0.04\; kpc^{-1}$. Another notable example is ultra-diffuse galaxy NGC 1052-DF2. It is believed to have roughly the same size as the Milky Way but has much less amount of dark matter than expected. The parameters in MOG have been estimated for this object in \cite{Moffat_2018} and found to be $\alpha = 1.30$ and $\mu =  0.443 \;kpc^{-1}$. For a weak gravitational field the phenomenological formulae for the parameters  $\alpha$ and $\mu$ can be obtained from \cite{Moffat_2008} and are given by
\begin{equation} \label{eq:8}
    \mu = \frac{D}{\sqrt{M}}
\end{equation}
\begin{equation} \label{eq:9}
    \alpha =  \alpha_{\infty}\frac{M}{\left( \sqrt{M}+E \right)^{2}}
\end{equation}
$D$ and $E$ appearing in the above equation are universal constants in MOG and take values  $D =6.25\times 10^{3} M_{\odot}^{1/2}kpc^{-1} $, $E = 2.5 \times 10^{4} M_{\odot}^{1/2}$  where $\alpha_{\infty}$ is defined as follows
\begin{equation} \label{eq:10}
    \alpha_{\infty} =  \frac{G_{\infty} - G_{N} }{G_{N}}
\end{equation}
where $G_{\infty}$ is the asymptotic limit of G for very large mass concentrations. To obtain $\alpha$ and $\mu$ for the Milky Way galaxy it is enough to substitute the mass of the Milky Way  $M_{MW} \approx 1.7 \times 10^{11}M_{\odot} $ and the value for $\alpha_{\infty} \approx 10$ gives $\alpha_{MW} = 8.89$.
In this paper, we will derive the MOG parameters by fitting the galaxy cluster data.

\section{Hydrostatic Mass Profiles from X-Ray Observations}

Clusters are the most massive gravitationally bound structures in the universe. In the current understanding of structure formation galaxy clusters are formed by the hierarchical sequence of mergers and accretion of smaller systems driven by gravity with dominant role of dark matter.
This causes the Intra Cluster Medium (ICM) to heat up to a temperature $T \sim (2-100)\times 10^{6}\;K$ and the electrons in the ICM radiate in the X-ray band via thermal bremsstrahlung.  Therefore one of the most robust methods to study their properties is based on  X-ray data. The success of such an approach lies in the ability of the modern instruments to spatially resolve gas temperature and density profiles which helps in the reconstruction of total mass of the cluster. In particular, measuring the X-ray surface brightness integrated along the line of sight provides the total gravitating mass. Considering the Inter Cluster Medium (ICM) matter to be a perfect gas  in hydrostatic equilibrium with  the gravitational potential of the cluster, one can use the acceleration experienced by the gravitating mass $g = GM_{tot}(<r)/r^{2}$ in the equation of hydrostatic equilibrium 
\begin{equation}\label{HSE}
    \frac{dP_{gas}}{dr} = -\rho(r)g(r)
\end{equation}
to obtain:
\begin{equation} \label{mtot}
    M_{tot}(<r) = \frac{r}{G \rho_{gas}(r)}\frac{dP_{gas}}{dr}
\end{equation}
where $ M_{tot}(<r)$ is the mass of the cluster measured within a given radius r. The assumption that ICM behaves as a perfect gas obeying the equation of state of the form $P_{gas}= n_{gas}k_{B}T$, leads to  
\begin{equation}\label{massprofile}
    M_{tot}(<r) =  -\frac{k_{B}}{\mu m_{\mu}}\frac{r T_{gas}(r)}{G} \left [ \frac{d\ln \rho_{gas}(r)}{d\ln r} +  \frac{d\ln T_{gas}(r)}{d\ln r}  \right ]
\end{equation}

It is convenient to describe galaxy cluster as a spherical region of a  radius $R_{\Delta}$ with mean density $\Delta$ times the critical density $\rho_{c,z}$ at the clusters redshift $z$ , where $\rho_{c,z} =  3H_{z}^{2}/8\pi G$ and $H_{z} =  H_{0}[ \Omega_{\Lambda}+\Omega_{m}(1+z)^{3}]^{0.5}$ is the expansion rate at the redshift $z$. 
Then the convenient quantity $M_{\Delta}$ is defined as 
\begin{equation}
    M_{\Delta} \equiv M_{tot}(< R_{\Delta}) =  \frac{4}{3}\pi \Delta \rho_{c,z} R_{\Delta}^{3}
\end{equation}
Similarly the total mass of the cluster within the virial radius $R_{\Delta}$ in MOG can be obtained by using the expression for the modified acceleration with the parameters $\alpha$ and $\mu$ given in Eq.(\ref{modifiedacceleration}). The most general expression for the total mass within the radius $r$ in STVG theory is: 
\begin{equation}\label{mogclustermass}
M_{STVG}(< r) = \frac{ M_{tot}(< r)}{1+\alpha-\alpha e^{-\mu r}(1+\mu r)},
\end{equation}
where $M_{tot}(<r)$ is the Newtonian mass  given by (\ref{mtot}). Hydrodynamical simulations predict that certain amount of energy content in the galaxy clusters could not yet be thermalized and could be present in the form of turbulence and bulk motions. Therefore masses estimated under the assumption that kinetic energy is fully thermalized might be biased and needed to be accounted for non-thermal pressure support to estimate total cluster mass. Although non-thermal pressure support is a difficult quantity to calculate yet there exist some promising approaches to this problem. For example, total baryon fraction can be used to estimate the integrated non-thermal pressure support \cite{Ghirardini_2017}. Another approach could be to use the Sunyaev-Zel'dovich effect. Its essence is that high energy ICM electrons change the temperature distribution of the Cosmic Microwave Background observed in the cluster direction through inverse Compton scattering: $ \frac{\Delta T_{CMB}}{T_{CMB}} = f(x)y$, where $y$ is the Compton parameter i.e. average fractional energy per collision multiplied by average number
of collisions (hence it is proportional to the integrated pressure), 
$f(x) = \left(x\frac{e^{x}+1}{e^{x}-1}-4 \right) \left(1+\Delta_{SZ}(x,T_{e})\right),$ with 
$x= h\nu/k_{B}T_{CMB}$ being the dimensionless photon frequency and  $\Delta_{SZ}(x,T_{e})$ the relativistic correction. 

In this paper we use the X-COP galaxy cluster sample \cite{X-COP_2}. The sample comprises 12 massive galaxy clusters (with redshifts $0.04 < z < 0.1$ ) observed in 
 X-rays on the XMM-Newton telescope in combination
with SZ effect observed within Planck all-sky survey. 
Planck SZ signal for these clusters has been recorded with the highest signal-to-noise ratio \cite{XCOP_SZ} translating into the especially good quality of $M_{tot}(< r)$ estimates with relative uncertainties around $5\%$. Hence, X-COP galaxy cluster sample provided a high-confidence total mass measurements. 

\section{Results and conclusions}

Based on the ideas outlined in previous section we estimated the the best fit values for the two parameters, characterizing the Moffat's STVG theory: $\alpha$ and $\mu$ for each galaxy from the X-COP sample. The fitting procedure was related to the chi-sqaure objective function: 
\begin{equation} \label{chisquare}
    \chi^{2} = \sum_{i=1}^{12}\left(  \frac{M^{th}_{STVG,i}(\alpha,\mu) - M^{obs}_{tot,i}}{\sigma_{M^{obs}_{tot,i}}}  \right)^{2}
\end{equation}
Reconstructed total masses $M^{obs}_{tot,i}$ of the clusters corrected for non-thermal support have been obtained from Table 2 of \cite{Eckert_2019}, and corresponding cluster redshifts and radii $R_{\Delta}$  can be found in Table 1 of \cite{Ettori_2019}. The best fitted parameters were obtained using MCMC approach using \textit{emcee} sampler  \cite{Foreman_Mackey_2013}. 

There is a subtlety, due to the system's mass dependence of $(\alpha, \mu)$ parameters, as discussed in Section 2, which makes the fitting not so straightforward. Our approach was the following. The $\mu$ parameter depends on mass of the system according to (\ref{eq:8}), where $D$ has universal meaning. Then, $\alpha$ also depends on mass according to (\ref{eq:9}), with $E$  being universal. $E$ and $D$ were fixed at the values obtained in \cite{Moffat_2008} (see Section 2). 
Consequently we used the expressions (\ref{eq:8}) and (\ref{eq:9}) in (\ref{mogclustermass}) to represent $M^{th}_{STVG,i}(\alpha,\mu)$ in (\ref{chisquare}). Accordingly, from the X-COP sample the $\alpha_{\infty}$ parameter has been fitted. The result is:  $\alpha _{\infty} = 9.12 \pm 2.72$ for $M_{200}$ data, and $\alpha _{\infty} = 9.99 \pm 3.40$ for $M_{500}$ data. Based on this fit one is able to assess the STVG parameters $(\alpha, \mu)$ representative of the X-COP sample. These results are shown in Table 1 and Table 2, where the values of $\alpha_{\Delta}$ and $\mu_{\Delta}$ obtained at $\Delta=(200,500)$ scales are reported for each cluster. The inverse variance weighted mean summarizes the values of $(\alpha, \mu)$ parameters at each overdensity scale $\Delta$: \begin{eqnarray} \nonumber
\alpha_{200,w.m} = 9.106 \pm 0.784; &&  \mu_{200,w.m.} 0.183 \pm 0.002 \\ \nonumber
\alpha_{500,w.m} = 9.970 \pm 0.979; &&  \mu_{500,w.m.} 0.238 \pm 0.003 
\end{eqnarray} 

In this paper we used the masses $M_{200}$ and $M_{500}$ measured within $R_{200}$ and $R_{500}$ radii in the X-COP sample of galaxy clusters to constrain the STVG modified gravity theory proposed by Moffat. 
Reliable mass measurements in X-COP sample was possible due to availability of good quality X-ray data together with measurements of Sunyaev-Zeldovich effect. ICM was assumed to be a perfect gas in hydrostatic equilibrium, which allowed to use the modified  gravitational potential in MOG to derive 
MOG modified masses. The modified  mass profiles is found to  depend on the parameter $\alpha$, a dimensionless parameter which measures the strength of gravitational interaction and $\mu$, the dynamical mass of the vector graviton. We  have shown that MOdified Gravity  can explain the mass of the X-COP galaxy cluster without non-baryonic matter. The estimated values of the parameters reported above, are  close to  the previous estimates \cite{rotationcurves2013} based on the analysis of galaxy rotation curves for the Milky Way galaxy and $\mu =  0.196 Mpc^{-1}$. In fact the $\alpha$ parameter is consistent with previous fits by other authors within the uncertainty range, while $\mu$ is more sensitive to the mass scale, as theoretically predicted. One can conclude that MOG is passing subsequent tests proving its consistency. Hence, MOG seems to be a promising alternative to General Relativity. Therefore it must be subjected to other stringent tests, e.g. related to gravitational lensing or gravitational waves. The presence of additional fields in MOG will give rise to extra polarizations in addition to $+$ and  $\times$ modes present in General Relativity. Future detectors such as LISA, Einstein Telescope, DECIGO could shed light in this path.

\section*{Acknowledgements}
We are grateful to John Moffat for helpful and stimulating discussions.

%

\bibliographystyle{unsrt}
\bibliography{ms.bib}

%


\clearpage

\begin{table} \label{Table2}
		\centering
		\caption{ Best fit values for  $\alpha_{200}$  and  $\mu_{200}$ estimated for X-COP cluster sample}
		\label{tab:example4}
		\begin{tabular}{lcccr} 
			\hline
			\hline
			Name & $M_{200}$ & $R_{200}$ & $\alpha_{200} $ & $\mu_{200}$ \\
			&($10^{14} M_{\odot}$) & $(Mpc)$ &  &$(Mpc^{-1})$ \\
			\hline
			A1644&  $6.58 \pm 0.66$ & $1.778 \pm 0.051$ & $9.102 \pm 2.715$  &  $0.244 \pm 0.012 $  \\ \\
			A1795& $ 6.76 \pm 0.36 $& $1.755 \pm 0.021$ & $9.102 \pm 2.715$ & $0.240 \pm 0.006 $\\ \\
			A2029&$13.29 \pm 0.69$ & $2.173 \pm 0.034$& $9.108 \pm 2.716 $ & $0.171 \pm 0.004$ \\ \\
			A2142& $ 16.37 \pm 0.89$ & $2.224 \pm 0.027$ & $9.109 \pm 2.717$ & $0.154 \pm 0.004 $  \\ \\
			A2255&  $10.70 \pm 0.68 $ & $2.033 \pm 0.081 $& $9.106 \pm 2.716 $& $0.191 \pm 0.006 $  \\ \\
			A2319&  $20.11 \pm 1.23$ & $2.040 \pm 0.035$ & $9.110 \pm 2.717$ &  $0.139 \pm 0.004 $ \\ \\
			A3158&  $7.34 \pm 0.41$  & $1.766 \pm 0.035$ & $9.103 \pm 2.715$& $0.231 \pm 0.006 $  \\ \\
			A3266&  $14.49 \pm 2.70$ & $2.325 \pm 0.074$& $9.108 \pm 2.716$ & $0.164 \pm 0.015  $ \\ \\ 
			A644 &  $8.35 \pm 0.61$  & $1.847 \pm 0.059$ & $9.104 \pm 2.715$ & $0.216\pm 0.008 $ \\ \\
			A85&  $9.56 \pm 0.49$  & $1.921 \pm 0.027$ & $9.105 \pm 2.716$& $0.202 \pm 0.005 $  \\ \\
			RXC1825&  $6.87 \pm 0.59$  & $1.719 \pm 0.024$ & $9.103 \pm 2.715$ & $0.238 \pm 0.010 $ \\ \\
			ZwC11215& $13.03 \pm 1.25$  & $2.200 \pm 0.069$ & $9.107 \pm 2.716$& $0.173 \pm 0.008 $ \\ \\
			\hline
		\end{tabular}
		\end{table}

\begin{table}\label{Table3}
\centering
\caption{ Best fit values for  $\alpha_{500}$  and  $\mu_{500}$ estimated for X-COP cluster sample}
\label{table2}
\begin{tabular}{lcccc} 
\hline
\hline
Name & $M_{500}$ & $R_{500}$ & $\alpha_{500} $ & $\mu_{500}$ \\
&($10^{14} M_{\odot}$) & $(Mpc)$ &  &$(Mpc^{-1})$ \\
\hline
			A1644&  $3.52\pm 0.21 $ &$ 1.054 \pm 0.020$ & $9.963\pm 3.391$  &  $0.333 \pm 0.01 $  \\ \\
			A1795&  $4.77 \pm 0.33$& $1.153 \pm 0.012$ & $9.967\pm3.392$ & $0.286\pm0.01 $\\ \\
			A2029& $8.98\pm 0.84$ & $1.423 \pm 0.019$ & $9.973 \pm 3.394 $ & $0.209\pm0.01$ \\ \\
			A2142&  $10.50\pm 0.73$ & $1.424 \pm 0.014$ & $9.975\pm3.395$ & $0.193\pm0.007 $  \\ \\
			A2255&  $5.87\pm 0.46$  & $1.196 \pm 0.026$ & $9.969\pm3.393$& $0.258 \pm0.01 $  \\ \\
			A2319&  $11.44\pm1.08 $ & $1.346 \pm0.017$ & $9.975\pm3.395$ &  $0.185\pm0.009 $ \\ \\
			A3158&  $4.53 \pm0.38 $  & $1.123 \pm 0.016$ & $9.967\pm3.392$& $0.294\pm0.012 $  \\ \\
			A3266&  $8.94\pm 0.57$ & $1.430 \pm 0.031$ & $9.973\pm3.394$ & $0.209\pm0.007  $ \\ \\ 
			A644 & $  6.03\pm 0.66$  & $1.230 \pm 0.035$ & $9.97\pm3.393$ & $0.255\pm 0.014$ \\ \\
			A85 &  $6.22\pm 0.49$ & $1.235 \pm 0.013$& $9.97\pm3.393$& $0.251\pm0.01 $  \\ \\
			RXC1825&  $3.94 \pm 0.32$  & $1.105 \pm 0.012$ & $9.965\pm3.391$ & $0.315\pm0.013 $ \\ \\
			ZwC11215& $7.67 \pm 0.53$ & $1.358 \pm 0.031$ & $9.972\pm3.3946$& $0.226\pm0.008 $ \\ \\
			\hline
\end{tabular}
\end{table}

\end{document}